# Mapping the Turn: An Eulerian Binormal-Axis Diagnostic for Recirculating 3D Flows

John Marshall Cooper*[1] and Wen Wu†[1]
*1. Department of Mechanical Engineering, University of Mississippi, University, MS, 38677, USA*

**Three-dimensional (3D) recirculating flows are often interpreted qualitatively from selected streamline visualizations. In separated flows, such recirculating motion is central to the drag modulation, but the local orientation of recirculation remains difficult to quantify in a field-based form. This work introduces an Eulerian binormal-axis diagnostic that locally evaluates the orientation of streamline turning at each point in the velocity field, yielding a spatially resolved field of the recirculating direction. Motivated by the Frenet–Serret binormal direction of a curved streamline, the diagnostic uses the velocity vector and its convective acceleration to extract the local streamline-turning axis without requiring explicit streamline integration. The resulting direction is encoded with barycentric RGB weights to visualize streamwise, spanwise, and wall-normal turning axis contributions. The diagnostic is first applied to Hill's spherical vortex, which provides a controlled analytic example of 3D recirculating motion for interpreting the binormal-axis direction and the associated barycentric RGB encoding. It is then applied to the mean field of a pressure-gradient-induced 3D separation bubble. The resulting visualizations show that the diagnostic reveals orientation changes that are not apparent from streamline visualization. The proposed diagnostic therefore converts qualitative streamline impressions into a spatially resolved measure of local streamline-turning orientation, providing a quantitative complement to conventional 3D flow visualization.**

## I. Introduction

Three-dimensional (3D) recirculating flows arise in a wide range of aerodynamic, turbomachinery, vehicle, and geophysical configurations. Such motions may occur as compact vortical structures, closed or nearly closed streamline patterns, or separated-flow recirculation regions, and they can strongly affect pressure loading, drag, heat transfer, mixing, and unsteady force generation. In aerodynamic applications, one important source of three-dimensional recirculation is flow separation. Unlike two-dimensional (2D) separation, 3D separation rarely admits a simple description in terms of a single separation point, a single reattachment point, or a closed recirculation bubble. Instead, the organization of the separating motion is set by the topology of the surface skin friction field, the associated separation and attachment lines, coherent vortical structures and three-dimensional transport pathways. The resulting flow structure is therefore intrinsically geometrical and fully three-dimensional [1, 2].

Even identifying separation and reattachment, which define the wall-bounded extent of a recirculating region, is nontrivial in 3D flows. In 2D steady flows, the classical picture of a wall-shear reversal provides a useful first description, but this interpretation does not directly generalize to unsteady or 3D configurations. Haller demonstrated that, in unsteady 2D flows, separation is not determined solely by instantaneous zero-skin-friction points, but by the Lagrangian evolution of near-wall material lines [3]. Subsequent extensions to 3D steady and unsteady flows further demonstrated that separation and attachment surfaces may originate from specific wall-based material lines whose identification requires a dynamical-systems description of the near-wall flow [4, 5]. Recent work has continued this Lagrangian perspective by using material-surface deformation and repulsion-rate measures to detect separation in laminar and turbulent separation bubbles [6]. These studies emphasize that separation detection is fundamentally tied to material detachment from the wall. However, wall-based separation information and selected streamline visualizations do not directly provide a field quantity for mapping the local orientation of recirculating motion. The present work therefore focuses on a complementary question: how can the local orientation of three-dimensional streamline turning be quantified in separating flows?

In practice, the spatial organization of a 3D recirculating region is still often interpreted through selected 3D streamlines. Such visualizations are valuable because they preserve the physical geometry of the motion, but they also have clear limitations. In a complex separation region, streamlines can be densely intertwined, and their dominant orientation may only become apparent through repeated changes of viewpoint, interactive rotation, or animation. In addition, streamline visualization is inherently dependent on the user-selected initiating locations: recirculation patterns that are not sampled by the prescribed seed points may be overlooked, even when they are dynamically important. Streamline plots are therefore primarily qualitative and selective: they reveal representative flow patterns, but they do not directly provide a field-based measure of how the local recirculating motion is oriented or how that orientation varies throughout the separating region.

*Ph.D. Student, AIAA Student Member
†Assistant Professor. AIAA Senior Member. Corresponding author. Email address: wu@olemiss.edu

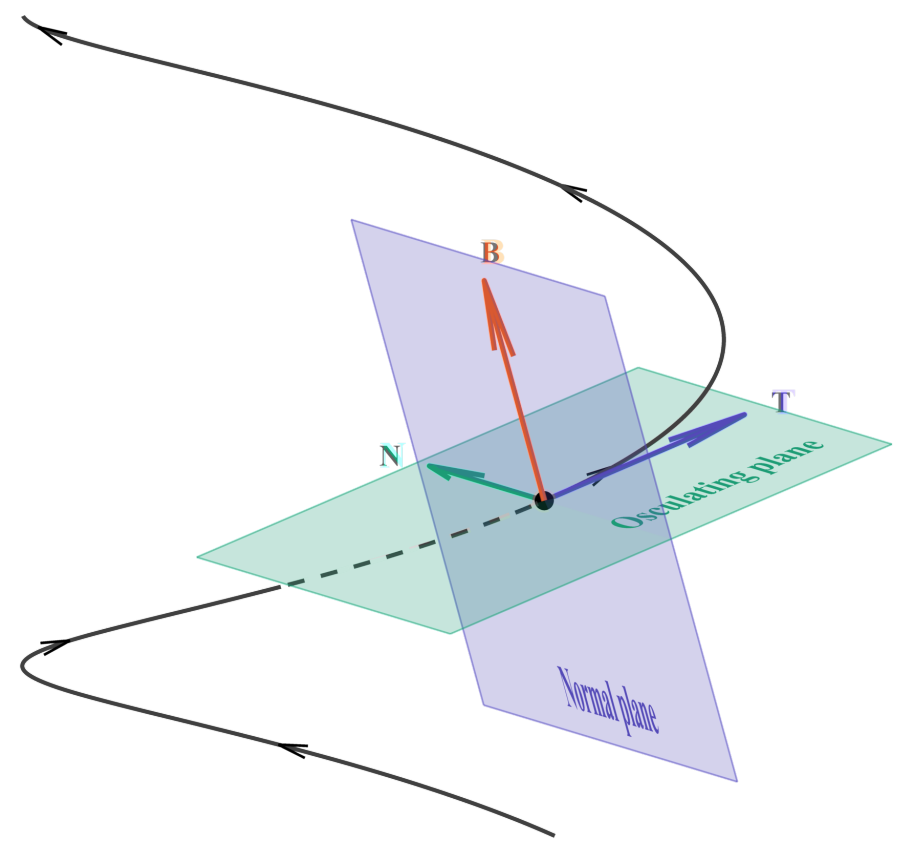

**Fig. 1 Frenet–Serret frame of a representative curved streamline. The tangent vector $\boldsymbol{T}$, principal normal vector $\boldsymbol{N}$, and binormal vector $\boldsymbol{B}$ define the local streamline geometry. The binormal vector is normal to the osculating plane spanned by $\boldsymbol{T}$ and $\boldsymbol{N}$, and corresponds to the local binormal-axis direction used in the present diagnostic.**

The present work introduces an Eulerian binormal-axis diagnostic for quantifying this local streamline-turning orientation directly from the velocity field. The diagnostic is constructed from the local velocity and its streamline-directional derivative, producing a binormal-axis direction that describes the orientation of local streamline bending. This converts the geometric information normally inferred from selected streamline plots into a field quantity that can be evaluated throughout a three-dimensional flow. It is therefore distinct from separation criteria, vortex-identification methods, and curve-by-curve Frenet–Serret analyses. In the application considered here, the diagnostic is evaluated to characterize the organization of recirculating motion in a pressure-gradient-induced 3D separation bubble. The result is a quantitative complement to conventional streamline visualization, mapping how local turning orientation of streamlines is organized in space.

## II. Eulerian Binormal-Axis Diagnostic

### A. Local Binormal-Axis Direction from the Eulerian Velocity Field

At any location where the local velocity is nonzero, we define the Eulerian directional derivative of the velocity field along the local streamline as

$$\mathbf{A} = (\mathbf{U} \cdot \nabla)\mathbf{U}. \tag{1}$$

Here, $\mathbf{A}$ is used as a geometric quantity that describes the local change of the velocity vector along the streamline. This definition does not require the flow field to be statistically averaged or steady. For an instantaneous velocity field, $\mathbf{A}$ should be interpreted as the directional derivative along the frozen-time instantaneous streamline.

The associated local curvature-axis vector is then defined as

$$\mathbf{C} = \mathbf{U} \times \mathbf{A}, \qquad \widehat{\mathbf{C}} = \frac{\mathbf{C}}{|\mathbf{C}|}, \tag{2}$$

where the unit vector is defined only when $|\mathbf{C}| \neq 0$.

The vector $\boldsymbol{C}$ defined here can be related to the binormal-axis direction of the Frenet–Serret frame of the local streamline spanned by the tangent ($\boldsymbol{T}$), principal normal ($\boldsymbol{N}$), and binormal ($\boldsymbol{B}$) unit vectors [7] (see figure 1). Streamline-aligned coordinate systems and moving-frame descriptions have a long history in fluid mechanics, particularly for interpreting curved shear flows and turbulent boundary layers. Recently, Finnigan [8] extended the streamline-coordinate formulation to three-dimensional turbulent flows and emphasized that the Frenet–Serret basis provides a natural moving frame in which the streamline tangent, principal normal, and binormal directions separate streamwise, curvature-normal, and binormal components of the dynamics. In that framework, however, the objective is to construct, when possible, a streamline coordinate representation of the governing equations. The present work uses the same local geometric interpretation of streamline curvature in a different way: rather than constructing a full coordinate system or integrating streamlines, we extract a local binormal axis directly from the Eulerian velocity field.

Specifically, writing local velocity as $\boldsymbol{U} = q\boldsymbol{T}$ with $q = |\boldsymbol{U}|$, it gives

$$\begin{aligned} \boldsymbol{A} &= (\boldsymbol{U} \cdot \nabla)(q\boldsymbol{T}) \\ &= q\frac{d(q\boldsymbol{T})}{ds} \\ &= \underbrace{q\frac{dq}{ds}\boldsymbol{T}}_{\text{Tangential Acceleration}} + \underbrace{q^2\kappa\boldsymbol{N}}_{\text{Normal/Centripetal Acceleration}}, \end{aligned} \tag{3}$$

where $s$ is arclength and $\kappa$ is the streamline curvature. Hence

$$\begin{aligned} \boldsymbol{U} \times \boldsymbol{A} &= q\boldsymbol{T} \times \left(q\frac{dq}{ds}\boldsymbol{T} + q^2\kappa\boldsymbol{N}\right) \\ &= q^2\frac{dq}{ds}\underbrace{\boldsymbol{T} \times \boldsymbol{T}}_{\text{zero}} + q^3\kappa(\boldsymbol{T} \times \boldsymbol{N}) \\ &= q^3\kappa\boldsymbol{B}. \end{aligned} \tag{4}$$

Thus, from a geometrical point of view, the tangential acceleration is eliminated by the cross product, leaving only the curvature-related term on the right-hand side, which quantifies the turning of the tangent vector. Thus, away from points of zero speed or zero curvature, where $|\mathbf{C}| \neq 0$,

$$\widehat{\mathbf{C}} = \mathbf{B}, \tag{5}$$

so that $\widehat{\mathbf{C}}$ represents the binormal direction of the local Frenet–Serret frame of the streamline. We therefore refer to $\widehat{\mathbf{C}}$ as the local binormal-axis direction in the remainder of the paper.

The corresponding non-negative binormal-axis directional weights are defined as

$$\boldsymbol{w} = \frac{\left(|\hat{C}_x|, |\hat{C}_z|, |\hat{C}_y|\right)}{|\hat{C}_x| + |\hat{C}_z| + |\hat{C}_y|}. \tag{6}$$

These weights encode the relative contributions of the fixed coordinate directions to the local binormal-axis direction. Since absolute values are used, they measure directional alignment but do not retain the sign of the binormal-axis orientation.

The essential distinction between the proposed Eulerian binormal-axis diagnostic and the classical Frenet–Serret description of a space curve lies in the choice of the primitive object. In the Frenet–Serret construction, the primitive object is a material or geometric curve, from which the tangent, normal, and binormal are obtained by differentiation with respect to arclength. In the present Eulerian binormal-axis diagnostic, the primitive object is the Eulerian velocity field. The quantity $\boldsymbol{A}$ is introduced because it is the directional derivative of the velocity along the local streamline and therefore contains the local turning information of the streamline without requiring explicit streamline integration. The vector $\boldsymbol{C}$ then gives the axis normal to the local osculating plane of that streamline. Thus, the method extracts the binormal-axis direction directly from the velocity-gradient field and subsequently projects it onto fixed coordinate directions; it is not a full Frenet–Serret frame construction, but an Eulerian diagnostic of local streamline-turning characterizing the recirculation orientation. The diagnostic is therefore streamline-based rather than pathline-based, except under conditions such as steady flow where streamlines and pathlines coincide.

### B. Directional Weights and Barycentric RGB Encoding

The weights defined in Eq. 6 provide a compact description of the orientation of the local binormal-axis direction, but they are not by themselves convenient for interpreting three-dimensional flow patterns. If the three weights are plotted as separate scalar fields, the dominant direction and its spatial variation must be reconstructed mentally from multiple panels. Similarly, if the weights are displayed in an abstract map detached from the physical domain, the directional information loses the geometric context that is essential for identifying organized streamline-turning patterns. Following the visualization philosophy of barycentric componentality maps, the present method transfers the weight information back into physical space by using the three directional weights directly as color-mixing coefficients.

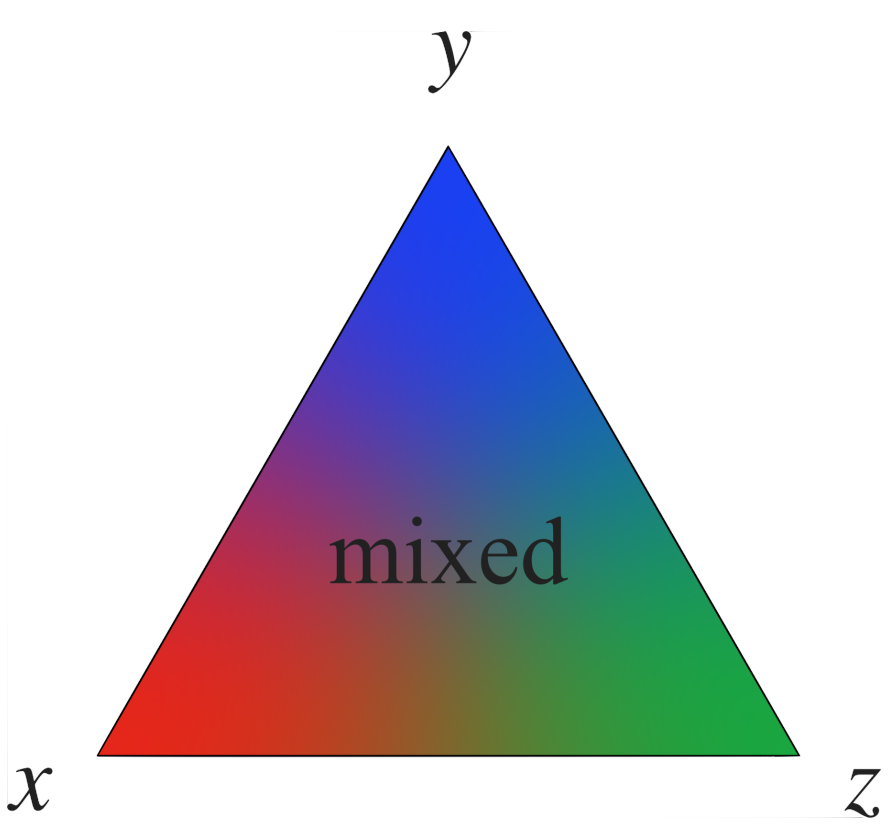

**Fig. 2 Barycentric RGB encoding of the binormal-axis directional weights. The red, green, and blue vertices correspond to streamwise $(x)$, spanwise $(z)$, and wall-normal $(y)$ binormal-axis contributions, respectively. Interior colors represent convex combinations of the three directional weights.**

To construct the color field, three base colors are assigned to the fixed coordinate directions,

$$\mathbf{C}_x = [0.90,\ 0.15,\ 0.10],\ \text{(i.e., red)} \tag{7}$$

$$\mathbf{C}_z = [0.10,\ 0.65,\ 0.25],\ \text{(i.e., green)} \tag{8}$$

$$\mathbf{C}_y = [0.10,\ 0.25,\ 0.95],\ \text{(i.e., blue)} \tag{9}$$

where the entries denote normalized RGB intensities. The color associated with the local binormal-axis direction is then obtained as the convex color combination

$$\mathbf{C}_B = w_x \mathbf{C}_x + w_z \mathbf{C}_z + w_y \mathbf{C}_y. \tag{10}$$

Because $w_x + w_y + w_z = 1$ and each weight is non-negative, the resulting color remains inside the triangular color space spanned by the three coordinate-direction colors (see figure 2). In this representation, a red-dominated color indicates that the local binormal-axis direction is primarily aligned with the $x$ direction, a blue-dominated color indicates dominance of the $y$ or wall-normal/vertical direction, and a green-dominated color indicates dominance of the $z$ or spanwise direction. Intermediate colors correspond to mixed orientations, rather than to discrete directional classes.

This color construction is barycentric in the sense that the displayed color is not chosen from a categorical lookup table, but is continuously interpolated from the three coordinate-axis contributions. Therefore, gradual changes in color along a streamline indicate gradual reorientation of the local binormal-axis direction, while abrupt color transitions identify regions where the streamline turning direction changes rapidly. When applied to the representative recirculating streamlines, the method preserves both pieces of information needed for interpretation: the streamline geometry remains visible in physical space, and the local binormal-axis direction is encoded directly on the same object. This avoids the loss of spatial context that would occur if the directional weights were examined only in a separate parameter space. The approach is particularly useful for distinguishing different recirculation orientations in three-dimensional separating flows. A streamline whose color remains nearly uniform is associated with a persistent orientation, whereas a streamline with alternating color segments undergoes a more complex three-dimensional reorientation.

More importantly, the primary utility of the present method is not limited to post-processing individually selected curves, as in a conventional Frenet–Serret analysis. Since the binormal-axis direction is obtained directly from the Eulerian velocity field through the local velocity and its Eulerian directional derivative, the directional weights can be evaluated at every location in the three-dimensional flow field without explicit streamline integration. The method therefore provides a field-based measure of local streamline-turning orientation; when evaluated for a 3D separating flow, it enables the spatial distribution of binormal-axis dominance within the recirculation region to be quantified directly.

### C. Relation to Vorticity, Velocity-Gradient-Based Diagnostics, and Lamb Vector

An intuitive alternative is to use the local vorticity direction to characterize the orientation of local turning or recirculating motion. This choice, however, is not equivalent to the binormal-axis direction considered here. Vorticity is determined by the antisymmetric part of the velocity-gradient tensor and may therefore reflect both local rotation and shear. More refined velocity-gradient-based diagnostics for rotation, including the $Q$ criterion, the $\lambda_2$ criterion, and the swirling strength $\lambda_{ci}$, are widely used to distinguish vortical motion from background strain or shear more robustly than vorticity alone. These diagnostics are well suited for characterizing the tensorial structure of $\nabla\mathbf{U}$ itself. By contrast, the present diagnostic is *geometric*: it is defined from the local streamline tangent and its convective acceleration, and identifies the binormal-axis direction associated with geometrical streamline bending. A recent decomposition of vorticity into orbital-rotation and spin contributions has also used the Frenet–Serret frame to interpret near-wall viscous-flow dynamics and boundary vorticity-flux mechanisms [9]. Although this orbital-rotation viewpoint is geometrically connected to the streamline-turning direction used here, the present diagnostic is not a decomposition of vorticity into rotational modes.

The relationship between the present binormal-axis vector and vorticity can be made explicit by using the vector identity

$$(\mathbf{U}\cdot\nabla)\mathbf{U} = \nabla\left(\frac{q^2}{2}\right) - \mathbf{U}\times\boldsymbol{\omega}, \tag{11}$$

where

$$\boldsymbol{\omega} = \nabla\times\mathbf{U}. \tag{12}$$

With $\mathbf{A} = (\mathbf{U}\cdot\nabla)\mathbf{U}$, the binormal-axis vector can therefore be written as

$$\mathbf{U}\times\mathbf{A} = \mathbf{U}\times\nabla\left(\frac{q^2}{2}\right) - \mathbf{U}\times(\mathbf{U}\times\boldsymbol{\omega}). \tag{13}$$

Using the triple-product identity gives

$$\mathbf{U}\times\mathbf{A} = \mathbf{U}\times\nabla\left(\frac{q^2}{2}\right) + q^2\boldsymbol{\omega} - \mathbf{U}(\mathbf{U}\cdot\boldsymbol{\omega}). \tag{14}$$

Equivalently, by defining the local unit tangent $\mathbf{T} = \mathbf{U}/|\mathbf{U}|$, this expression becomes

$$\mathbf{U}\times\mathbf{A} = \mathbf{U}\times\nabla\left(\frac{q^2}{2}\right) + q^2\left[\boldsymbol{\omega} - (\boldsymbol{\omega}\cdot\mathbf{T})\mathbf{T}\right]. \tag{15}$$

This form shows that the binormal-axis vector used in the proposed diagnostic is related to, but distinct from, the vorticity vector. The second term retains only the component of vorticity perpendicular to the local velocity direction, while the streamwise component of vorticity is removed. In addition, the first term introduces the contribution from spatial variations of kinetic energy, or equivalently from gradients of the velocity magnitude. Thus, the local binormal-axis vector is not simply a rescaled vorticity vector. It combines the cross-stream vorticity contribution with the speed-gradient contribution that is required to describe the turning of the streamline tangent.

The present binormal-axis vector is also connected to the Lamb vector. With the Lamb vector defined as $\boldsymbol{L} = \boldsymbol{\omega}\times\boldsymbol{U}$, Eq. 13 can also be written as

$$\boldsymbol{U}\times\boldsymbol{A} = \boldsymbol{U}\times\nabla\left(\frac{q^2}{2}\right) + \boldsymbol{U}\times\boldsymbol{L}.$$

Thus, the binormal-axis vector can be interpreted as the cross product between the velocity vector and the sum of the kinetic-energy gradient and the Lamb vector.

## III. Preliminary Results

### A. Hill's Spherical Vortex

As a controlled analytic application, we first consider Hill's spherical vortex, a classical exact solution of the Euler equations that represents compact three-dimensional recirculating motion [10]. Its closed streamlines inside a spherical domain provide a simple setting for interpreting the physical meaning of the binormal-axis diagnostic. The purpose of this case is to demonstrate how the proposed barycentric RGB encoding, based on the binormal-axis directional weights, represents the local turning-axis organization of a known recirculating flow.

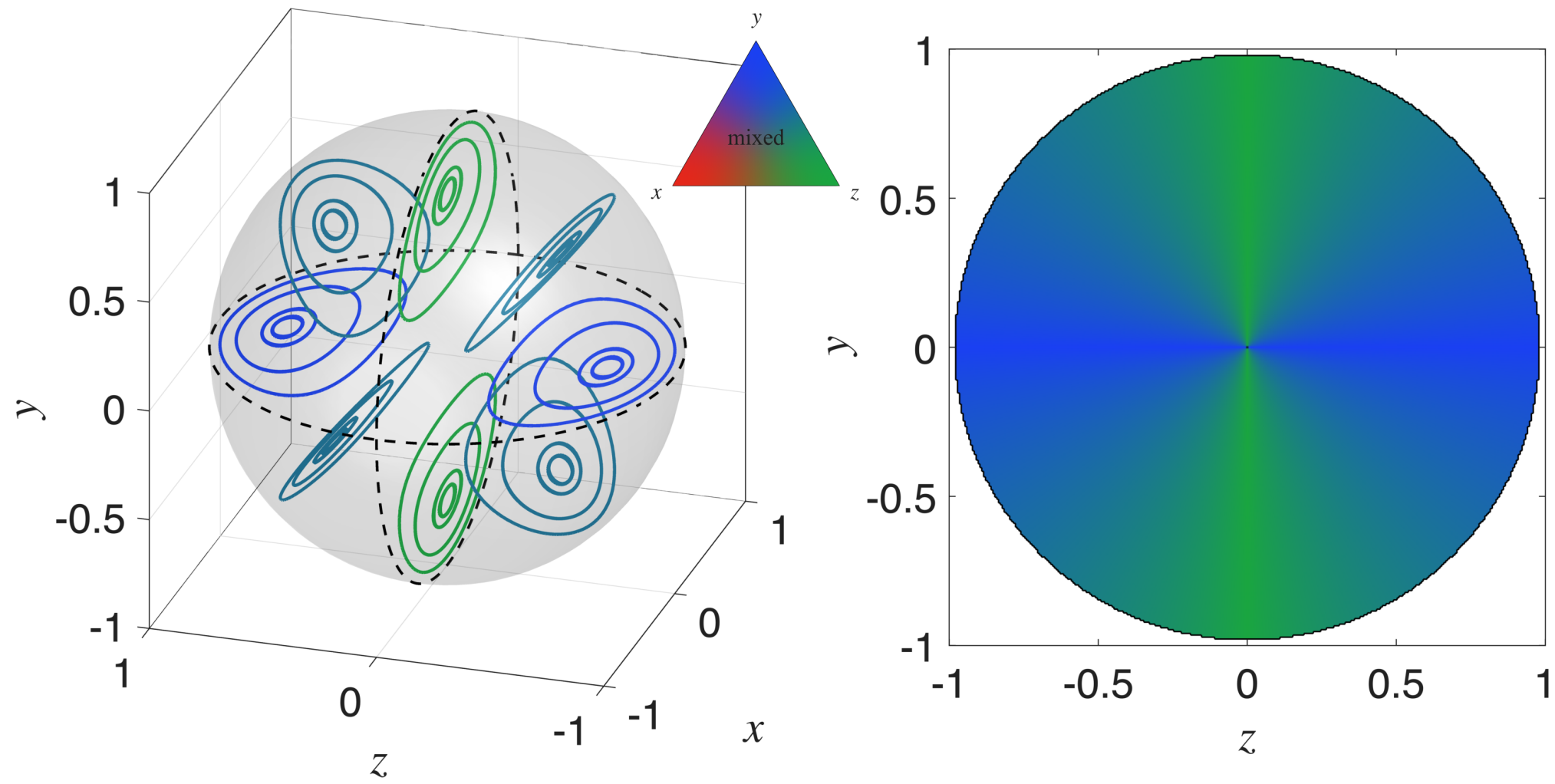

**Fig. 3 Application of the Eulerian binormal-axis diagnostic to Hill's spherical vortex. The left panel shows representative streamlines inside the spherical domain, colored by the barycentric RGB encoding of the local binormal-axis direction. The transparent gray sphere marks the outer boundary of the analytic recirculating flow. The right panel shows the corresponding binormal-axis color field on an $x$-normal slice through the center of the sphere. The color field reflects the circumferential binormal-axis direction of the axisymmetric recirculating flow.**

We take the symmetry axis of Hill's spherical vortex to be aligned with the $x$ direction. Using the axisymmetric coordinates $(x, \rho, \phi)$, where $\rho = \sqrt{y^2 + z^2}$, the internal velocity field is written as [10, 11]

$$\mathbf{U} = U_x \mathbf{e}_x + U_\rho \mathbf{e}_\rho, \qquad U_\phi = 0, \tag{16}$$

with

$$U_x = \frac{3}{2} U_0 \left[ 1 - \frac{x^2 + 2\rho^2}{R^2} \right], \quad U_\rho = \frac{3}{2} U_0 \frac{x\rho}{R^2}. \tag{17}$$

Here $R$ is the sphere radius and $U_0$ sets the velocity scale. The flow is axisymmetric and swirl-free, so each streamline remains in a meridional plane spanned by $\mathbf{e}_x$ and $\mathbf{e}_\rho$. Two representative intersections of such meridional planes with the spherical boundary are shown by the dashed reference curves in Fig. 3. Consequently, the local osculating plane of each streamline is the corresponding meridional plane, and the binormal-axis direction is circumferential about the $x$ axis,

$$\mathbf{B}_{\text{exact}} = \pm \mathbf{e}_\phi. \tag{18}$$

In Cartesian components,

$$\mathbf{e}_\rho = \left(0, \frac{y}{\rho}, \frac{z}{\rho}\right), \qquad \mathbf{e}_\phi = \mathbf{e}_x \times \mathbf{e}_\rho = \left(0, -\frac{z}{\rho}, \frac{y}{\rho}\right). \tag{19}$$

Thus, up to an arbitrary sign,

$$\mathbf{B}_{\text{exact}} = \left(0, -\frac{z}{\rho}, \frac{y}{\rho}\right) \text{ when } \rho \neq 0. \tag{20}$$

The sign ambiguity is immaterial for the barycentric RGB encoding because the directional weights are defined from the absolute values of the binormal-axis components.

Figure 3 shows the result of applying the diagnostic to this analytic recirculating flow. The streamline visualization in the left panel provides the geometric context of the closed recirculating motion inside the spherical domain, while the barycentric coloring overlays the local binormal-axis orientation on the same curves. The color variation therefore makes the azimuthal organization of the local turning-axis direction visible directly on the streamline geometry, rather than requiring the orientation to be inferred from the curve shape alone.

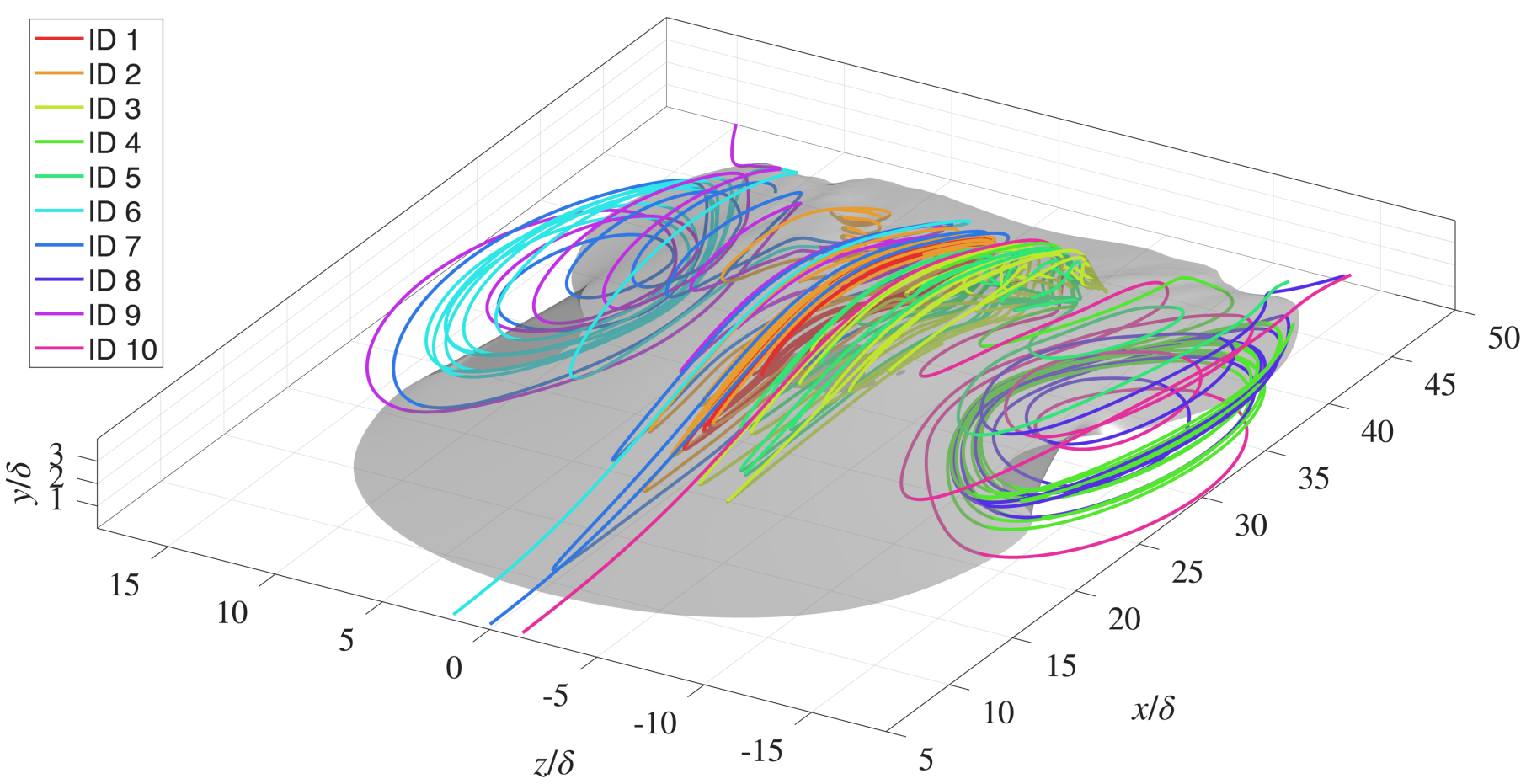

**Fig. 4 Representative mean streamlines used for recirculation-orientation analysis. The gray isosurface denotes the boundary of the reversed-flow region, defined by $U = 0$. The streamline colors in this figure are used only to distinguish streamline indices and do not represent the barycentric RGB encoding.**

The $x$-normal slice in the right panel shows the corresponding field representation of the same orientation information. Rather than depending on selected streamline seeds, the diagnostic assigns a local turning-axis orientation throughout the spherical cross-section. The resulting color field therefore gives a spatial view of how the local streamline-turning orientation is organized around the symmetry axis. This simple analytic example serves as a bridge between streamline-based visualization and the field-based application to the three-dimensional separation bubble considered next.

### B. Pressure-Gradient-Induced Three-Dimensional Separation Bubble

The diagnostic is applied to the mean flow of case z10v85 from the pressure-gradient-induced three-dimensional laminar separation bubble DNS reported in Ref. [12]. In that configuration, an incompressible laminar boundary layer at $Re = U_\infty \delta / \nu = 1,000$ develops over a flat plate, while a prescribed suction-and-blowing distribution at the top boundary imposes an adverse/favorable pressure-gradient sequence. The streamwise suction-and-blowing profile is modulated in the spanwise direction by a Gaussian weighting function, producing a localized three-dimensional pressure-gradient field without introducing geometric curvature of the wall. The case designation z10v85 corresponds to a spanwise $1/e$ damping half-width $z_0 = 10\delta$ and peak suction/blowing magnitude $V_0 = 0.85U_\infty$. This case forms a finite-width three-dimensional mean recirculation region with a streamwise length of $L_{\text{sep}} = 37\delta$, a minimum spanwise width of $W_{\text{sep,min}} = 19\delta$, and a maximum spanwise width of $W_{\text{sep,max}} = 31\delta$. It is therefore a useful test case for the present diagnostic because the recirculating motion is strongly three-dimensional and spatially localized, but still generated over a nominally flat wall by a controlled pressure-gradient distribution rather than by surface geometry.

The three-dimensional character of the mean recirculation region is first evident from the representative streamlines shown in Fig. 4. The gray isosurface marks the boundary of the reversed-flow region, defined by $U = 0$. Each selected streamline contains a segment that lies within this finite-width reversed-flow region and exhibits closed or nearly closed looping motion, indicating the presence of a three-dimensional recirculation bubble. These streamlines provide a direct geometric view of the recirculating motion, but they also illustrate the limitation of streamline visualization alone: because the curves are intertwined and their apparent orientation depends strongly on the viewing angle, the dominant axis about which the local motion turns is difficult to identify from the streamline geometry alone.

This limitation of visual inspection of three-dimensional streamlines can be mitigated by the proposed method. Figure 5 shows the same representative streamlines colored using the barycentric RGB encoding of the binormal-axis directional weights. The color variation along each curve demonstrates that the local binormal-axis direction is non-uniform along individual three-dimensional streamlines. Some portions of the streamlines are dominated by a spanwise-oriented binormal axis (green) or a wall-normal-oriented binormal axis (blue), whereas other portions contain mixed streamwise, wall-normal, and spanwise contributions. Thus, the recirculating motion is quantified through the local binormal-axis direction, whose variation along the streamlines reflects the intrinsically three-dimensional organization of the separating flow.

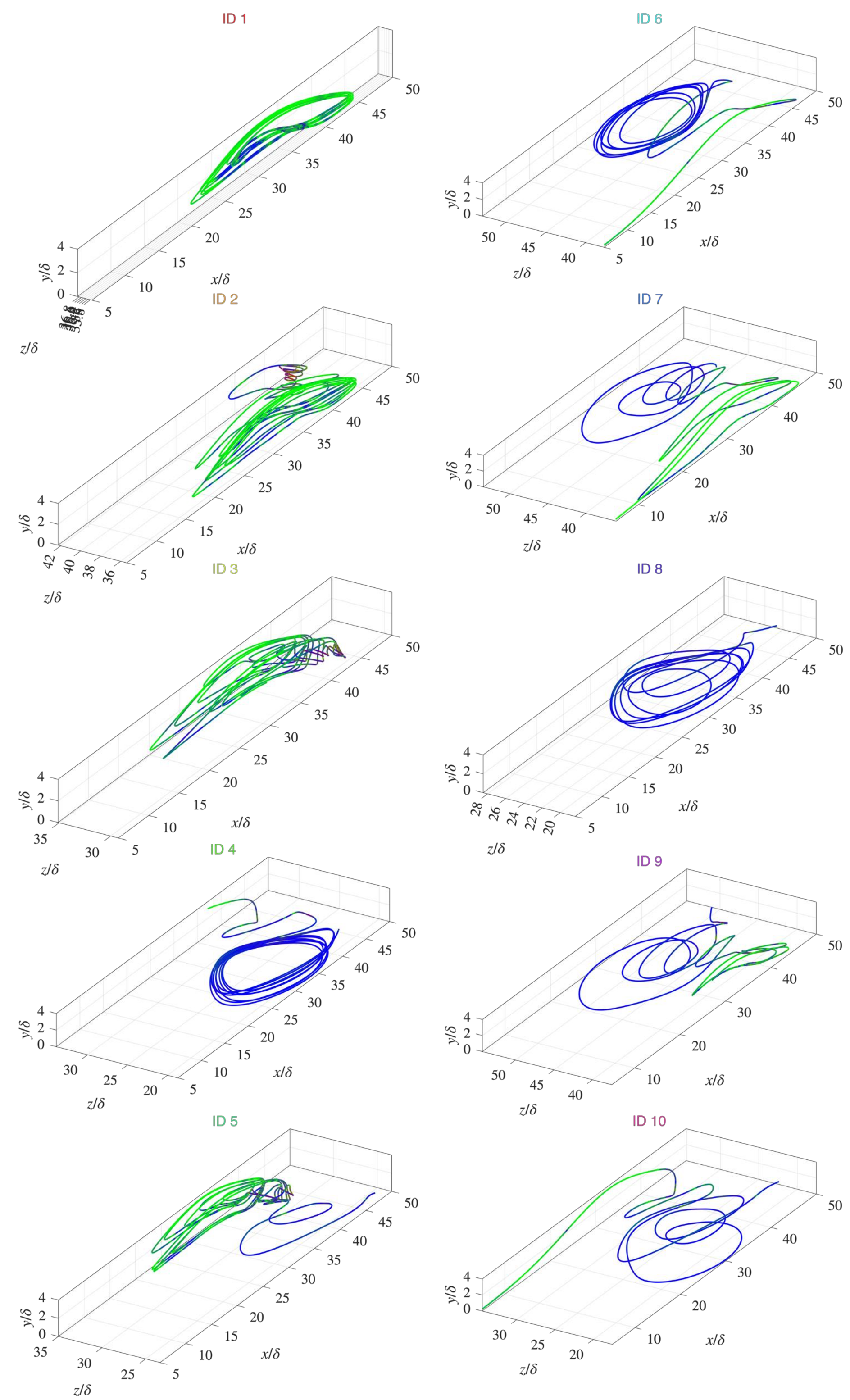

**Fig. 5 Barycentric RGB encoding of the binormal-axis directional weights along the representative mean streamlines. Each streamline is shown in a separate panel, with colors assigned from the local directional weights. The numerical label identifies the corresponding streamline in Fig. 4.**

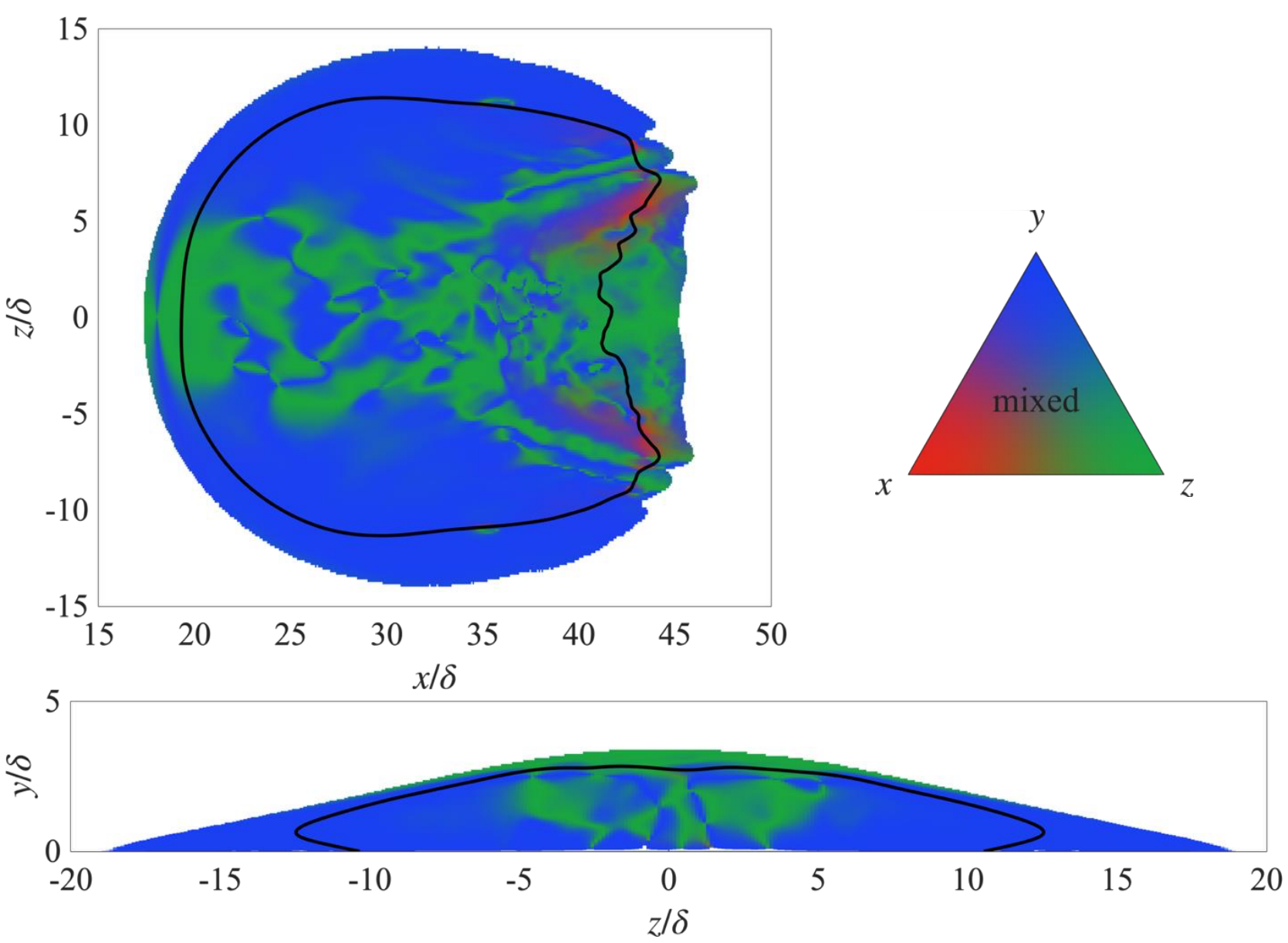

**Fig. 6 Contours of the binormal-axis direction for case z10v85. The upper-left panel shows a wall-parallel plane at $y/\delta = 1.22$, and the lower panel shows a transverse plane at $x/\delta = 28$. The color triangle indicates the barycentric RGB encoding of the streamwise, spanwise, and wall-normal binormal-axis contributions. The black contour denotes $U = 0$. Only points inside the recirculation region are shown.**

The advantage of formulating the diagnostic as an Eulerian field quantity is demonstrated more clearly in Fig. 6. Instead of inferring the organization of the recirculating motion from user-selected streamline seeds, the binormal-axis direction is evaluated throughout the recirculation region and displayed on a wall-parallel plane at $y/\delta = 1.22$ and on a transverse plane at $x/\delta = 28$. The black contour denotes $U = 0$, and only points inside the recirculation region are shown. The resulting field reveals that the portion of the bubble associated with conventional streamwise–wall-normal recirculation, i.e., a primarily spanwise binormal axis (green), is concentrated near the midspan region, approximately within $|z|/\delta \lesssim 5$. Away from this central region, much of the recirculation region is instead dominated by a wall-normal binormal axis (blue), indicating recirculating motion organized primarily in the streamwise–spanwise plane. The transverse view further shows the wall-normal distribution of the midspan recirculation mode: the spanwise-axis contribution is strongest in the central upper portion of the bubble, while the surrounding regions transition to mixed and wall-normal-axis orientations.

In addition, near the downstream reattachment side of the bubble, the diagnostic identifies distinct streamwise-axis-dominated regions (red) near the two spanwise edges of the reattaching zone. A streamwise binormal axis corresponds to local turning in cross-stream/transverse planes and is consistent with the formation of strongly three-dimensional reattachment motions along the two sides of the wake. This observation provides a field-based explanation for the dual-wake behavior reported in Ref. [12]: the two turbulent wake branches are not merely downstream consequences of the separating shear layer, but are connected to localized changes in the binormal-axis orientation of recirculating motion near the spanwise edges of the mean bubble.

The spatial distribution of these distinct streamline-turning patterns within the recirculation region would be difficult to quantify robustly from a small number of streamlines, because it depends on the local binormal-axis orientation of streamline turning rather than on the overall shape of any single streamline. This illustrates the primary utility of the present method: it converts the qualitative impression of three-dimensional recirculation into a spatially resolved map of the dominant local binormal-axis direction.

The present results also indicate that improved statistical convergence and a smoother representation of the three-dimensional recirculation region are needed for robust binormal-axis identification, because the diagnostic depends on higher-order velocity-gradient quantities. These issues will be addressed in the full paper, together with applications to additional three-dimensional separating flows.

## Acknowledgments

The authors acknowledge the support from NSF Grant No. 2439750, monitored by Dr. Ronald D. Joslin.